\documentclass[submission,copyright,sharealike]{eptcs}
\providecommand{\event}{TFPIE 2022} 
\usepackage{graphicx}
\usepackage{overpic}
\usepackage{listings}
\usepackage{fancyvrb}
\usepackage[shortlabels]{enumitem}
\usepackage{amsmath}
\usepackage{amssymb}
\usepackage{amstext}
\usepackage{amsthm}

\title{Teaching Interaction using State Diagrams}
\author{Padma Pasupathi
\institute{McMaster University, Hamilton, Ontario}
\email{pasupatp@mcmaster.ca}
\and
Christopher W. Schankula
\institute{McMaster University, Hamilton, Ontario}
\email{schankuc@mcmaster.ca}
\and
Nicole DiVincenzo
\institute{McMaster University, Hamilton, Ontario}
\email{n.divincenzo@hotmail.com}
\and
Sarah Coker
\institute{McMaster University, Hamilton, Ontario}
\email{cokers@mcmaster.ca}
\and
Christopher Kumar Anand
\institute{McMaster University, Hamilton, Ontario}
\email{anandc@mcmaster.ca}
}
\def\titlerunning{Teaching Interaction}
\def\authorrunning{Pasupathi, Schankula, DiVincenzo, Coker, and Anand}

\begin{document}
\maketitle

\begin{abstract}
To make computational thinking appealing to young learners, initial programming instruction looks very different now than a decade ago, with increasing use of graphics and robots both real and virtual.  
After the first steps, children want to create interactive programs, and they need a model for this.  
State diagrams provide such a model.

This paper documents the design and implementation of a Model-Driven Engineering tool called SD~Draw,
that allows even primary-aged children to draw and understand state diagrams,
and create modifiable app templates in the Elm programming language using the model-view-update pattern standard in Elm programs.
We have tested this with grade 4 and 5 students.
In our initial test, we discovered that children quickly understand the motivation and use of state diagrams using this tool,
and will independently discover abstract states even if they are only taught to model using concrete states.
To determine whether this approach is appropriate for children of this age we wanted to know:
do children understand state diagrams, do they understand the role of reachability, and are they engaged by them?
We found that they are able to translate between different representations of state diagrams, 
strongly indicating that they do understand them.
We found with confidence $p=0.001$ that they do understand reachability by refuting the null hypothesis that they are creating diagrams randomly.
And we found that they were engaged by the concept, with many students continuing to develop their diagrams on their own time after school and on the weekend.\end{abstract}


\section{Introduction}

McMaster Start Coding (\url{http://outreach.mcmaster.ca}) has introduced 25,000 children in Grades 4 to 8 to functional programming in Elm over the last 5 years using socially constructive learning.  
During the pandemic, we started offering on-line camps with different themes, 
including camps focussed on interactive applications.
Using a human-centered design approach, we interviewed past instructors to identify gaps and opportunities.
We identified the poor translation to the virtual environment of whiteboard state diagrams and poor functioning of the noisy state diagram game we played in classrooms as gaps,
and a better version of a Model-Driven Engineering (MDE) tool as an opportunity.
Our program is built on lessons, and we have videos\footnote{https://www.youtube.com/c/McMasterStartCoding/videos} for self-learners and for use in training mentors.  
Lesson 8 covers state diagrams.
Pre-pandemic, after teaching using our state-diagram game using a (physical) white board, we helped children use PAL~Draw \cite{schankula2020newyouthhack}, 
an MDE tool originally developed for advanced developers who could manipulate Petri nets with embedded state diagrams.
Although it is overly complex, children seemed to enjoy working with it in person, but it was just too awkward to use virtually.
Translating maps children draw into code is difficult for beginning programmers, but there is no reasonable alternative mathematical structure for describing interaction in simple ``adventure'' games.

Inspired by the Event-Driven Programming (EDP) and MDE literature,
we decided to make a better tool for state diagramming
with support for code generation.
The new tool adheres better to Norman's principles\footnote{For a discussion of these principles and an analysis of the new app see \cite{pasupathi2021sd}.},
generates complete programs (with working buttons for all transitions),
and went through many iterations internally before we tried to use it with children from Grades 4 and 5.
Most of our outreach activities are time-constrained,
so our initial test was restricted to 3.5 hours per class of virtual instruction, followed by one hour of challenges a week later.
Surprisingly,
we could answer affirmatively all of our questions about student understanding in this short intervention,
including statistically significant results on their understanding of reachability,
and the observation that they could both translate between different representations of state diagrams and spontaneously use abstract states like ``DragonIsDead'' or ``GameOver'' versus concrete states like ``Park'' or
``Mountain'' in their diagrams.

\smallskip
We also wanted to think about the place an improved tool could have in our outreach efforts,
which can be summarized by the following research questions: 
\begin{enumerate}
    \item[RQ1] Do grade 4-5 students demonstrate an understanding of State Diagrams by being able to translate between different representations? \item[RQ2] Do grade 4-5 students demonstrate equal facility for translating between different representations of state diagrams? 
    \item[RQ3] Can grade 4-5 students understand the role of reachability?
    \item[RQ4] Are grade 4-5 students engaged by state diagrams and their applications to adventure games?
    \item[RQ5] Do grade 4-5 students understand abstract and concrete states equally well? We define \textit{concrete} to be physical places as opposed to \textit{abstract} states such as ``being free''.  Will students presented with concrete states generalize to abstract states without prompting? 
\end{enumerate}
Operationally, a concrete state is a place a
child would draw on a map or blueprint.
Although this necessarily depends on the 
developmental level of the child,
the researchers found it easy to classify the states created by children.

The remaining sections will explain our \textit{methods} for this design iteration, including app design, curriculum design and evaluation through observation and challenges; \textit{results} of the challenges and analysis of the children's state diagrams; \textit{discussion} of the results leading to proposed \textit{future work}.

\section{Background}

In this section we provide background of our program as well as relevant research on
which we built our tool and instruction.

\subsection {McMaster Start Coding Program}
Our McMaster University Outreach Program has been operating for the past decade. 
A mainly  volunteer group of undergraduate and graduate students develop lesson plans and deliver 
free workshops to schools, public libraries, and community centres in the 
Hamilton, Ontario, Canada area \cite{o2017code}. During the COVID-19 pandemic, the program has shifted online
and has taught a record number of students. Since 2016, we have taught over 26,000 students in nearly 1,000 
classrooms. The goal of the program is to foster interest and ability in STEM subjects through coding, 
especially for those groups who are underrepresented in STEM subjects, such as girls and underprivileged
youth.

To support these workshops, we have developed several tools, including:
\begin{enumerate}
\item An open-source Elm graphics library, GraphicSVG\footnote{\url{https://package.elm-lang.org/packages/MacCASOutreach/graphicsvg}}.
\item An online mentorship and Elm compilation system incorporating massive collaborative programming tasks, including the Wordathon\footnote{http://outreach.mcmaster.ca/\#wordathon2019} and comic book storytelling\footnote{http://outreach.mcmaster.ca/\#comics2019}.
\item A curriculum for introducing graphics programming designed to prepare children for algebra \cite{EPTCS270.2}.
\item A type- and syntax-error-free projectional iPad Elm editor, ElmJr\footnote{\url{https://apps.apple.com/ca/app/elmjr/id1335011478}}.
\end{enumerate}

\subsection{Functional Programming}
Functional programming \cite{krishnamurthi2019programming} is a value-oriented programming paradigm, consisting of functions. Functions consume and produce values.
There are no loops, and conditional expressions replace conditional statements,
but functions are first-class values and can, e.g., be passed as parameters.
There are two variations in functional programming languages: (1) typed or not and (2) eager or lazy. (Some researchers would also include pure versus impure as variations.)
These variations lead to differences in programming style. 

Many non-functional programming languages are adopting functional features, 
including Scala, Swift and Python.

Krishnamurthi et al \cite{krishnamurthi2019programming} agree with the common perception that writing programs in imperative programming languages is much easier as the state provides convenient communication channels between parts of a program, but this makes reasoning and debugging harder, whereas on the other hand functional programming has the opposite affordances. Students studying imperative (and object-oriented) programming are taught different skills and programming styles which reveal that the way of approaching programming and problem-solving differs in students studying different paradigms. Functional programming students perform better by having high level structures and composing solutions out of simpler functions than object-oriented students who try solving the entire problem in a single traversal of data. Functional programming students create short functions for specific tasks, which create intermediate data. They also use \texttt{filter} and \texttt{map} rather than loops and non-general library functions. Thus, we should expect that a student who learns Java after learning functional programming may well program with different patterns than a student whose prior experience was entirely imperative.

Prior to teaching with Elm, we used Alice, Haskell+Gloss, Objective-C and Python 
as a minor component in our grade 4-8 outreach program composed mostly of non-keyboard activities, like inventing an algorithm and racing to reconstruct the inside of a dog's stomach from tomography data,
or encoding and decoding binary images transmitted between iPads. 
Since Elm had a no-install web environment, and looked a bit like Gloss,
we tried it as a way of taking programming out of our labs and into classrooms.
It was such a success, we were able to dispence with all of our classroom management hacks, and could focus on teaching!

\begin{figure*}[ht]
    \centering
    \includegraphics[width=0.9\textwidth]{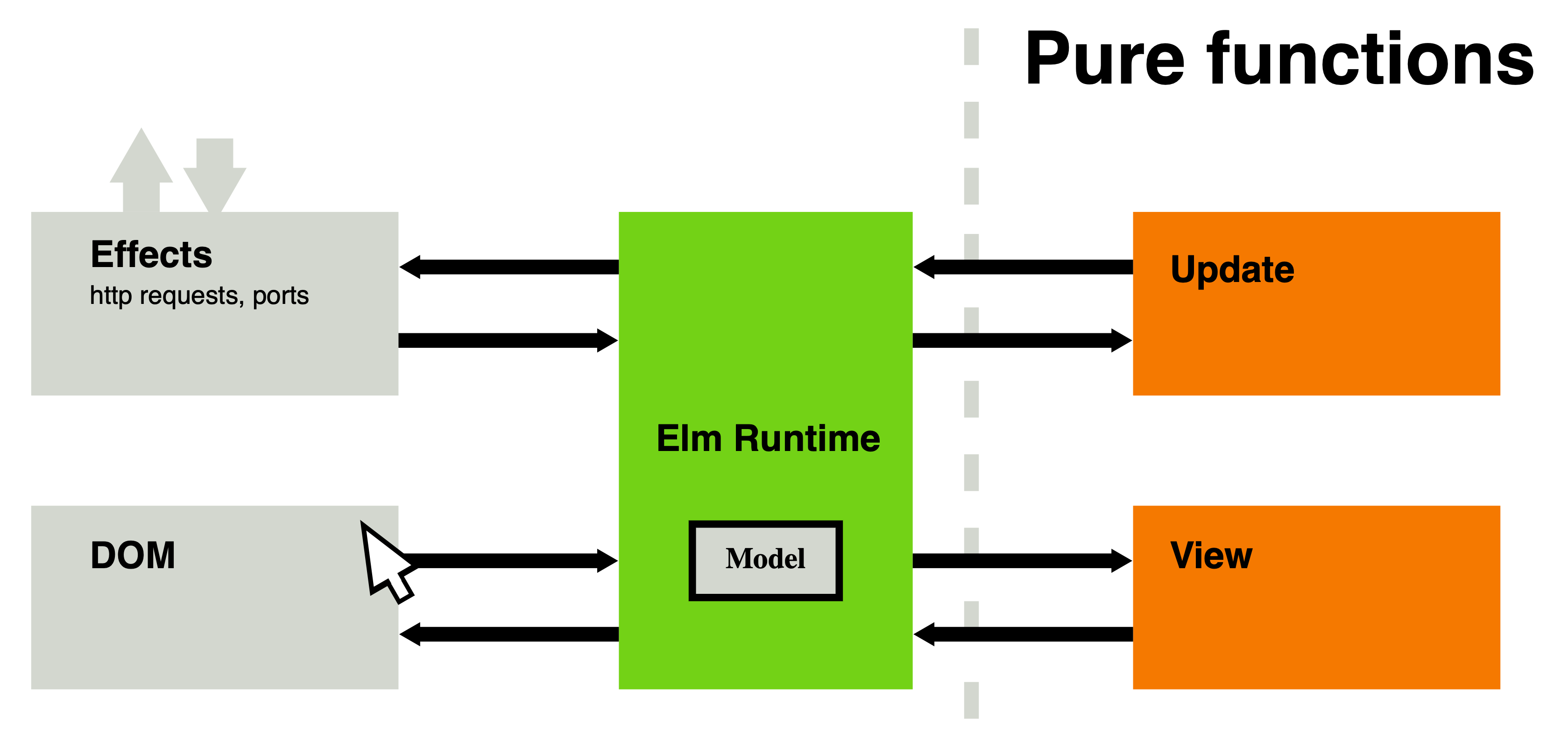}
    \caption{The Elm Architecture is built on the model-view-update pattern.
       User code to the right consists of pure functions which are called 
       by the run-time system to either render the current state, 
       represented as a \texttt{Model} data structure, or to calculate an $update:$ \texttt{Message $\to$ Model $\to$ Model}. 
       An animation of the resulting dataflow is presented in the slides
       \url{https://macoutreach.rocks/TFPIE2019Slides.html}.
       }
    \label{fig:TheElmArchitecture}
\end{figure*}

\subsection{Elm Language}
Elm (\url{https://elm-lang.org}) is a functional language designed for the development of front-end web applications \cite{czaplicki2012elm},
and promoted to front-end developers as a way of avoiding the many 
software quality issues which plague JavaScript programs.
Its grammar and concrete syntax, based on Haskell,
is intentionally simple. For example, it has no support for user-defined type classes. 
In addition to strictly enforcing types, the Elm compiler also forces programmers to 
follow best practices,
such as disallowing incomplete case coverage in case expressions.
Elm apps use a model-view-update pattern in which users write pure functions
and the run-time system handles side effects.
Elm code compiles to JavaScript simplifying deployment 
and visualization.
Figure~\ref{fig:TheElmArchitecture} shows how the Elm run-time system
mediates between the (stateful) world and the two pure functions
\begin{align}
update &: \text{\texttt{Msg}} \to \text{\texttt{Model}} \to \text{\texttt{Model}} \\
view &: \text{\texttt{Model}} \to \text{\texttt{Html}}
\end{align}
involving the user-defined algebraic data types  \texttt{Msg} and \texttt{Model}.
The main advantage of this pattern is that 
the key aspects of the interaction are represented compactly as single data types (for messages/events and state) or functions (state update and view).
This makes it easy to present multiple views of the same state.
When programming discrete interactions, programmers often make two bookkeeping errors:
they forget to allow an action, e.g., by omitting a button, 
or they forget to handle the result of a user action. 
The first error is still possible in Elm,
but the second is not possible once the programmer adds a message for the interaction
to the \texttt{Msg} type,
because the compiler analyzes case expressions and flags unhandled cases as errors.

While many practitioners at trade and generalist computer science conferences 
have objected to us that functional programming should be reserved for expert users, many of the features 
useful for experts (strict types, pure functions, well-thought-out error messages) are very useful for beginners. 
Although our youngest learners do not have a clear model for functions,
it is essential to their future success in high-school mathematics that they gain a correct mental model of 
a function as 
``something which takes an input and gives an output'',
which matches Elm semantics, unlike most imperative and object-oriented languages.

\subsection{Algebraic Thinking}

New approaches to teaching ``computational thinking'' to younger and younger children are well accepted 
in the schools we visit.
It is now common for children to see robots as early as kindergarten,
and the idea that a program 
is a set of instructions like a recipe
is understood by the teachers we visit.

Our approach is complementary to the recipe approach,
and designed with the primary goal to 
prepare children for learning algebra \cite{EPTCS270.2}.
In summary,
our graphics library contains a
large number of functions which correspond
to the shapes and transformations in the
math curriculum.
Each function has a small number of arguments,
making them easy to learn,
and their last input and output types are
often \texttt{Shape} or \texttt{Stencil},
so that they lend themselves to composition
(as well as to use in higher-order functions).
It is possible for children to create interesting graphics in their first hour of coding,
and within the second hour, 
we introduce variables as a time-saving feature,
and user-defined functions in the next hour as a an even more powerful time-saver.

As educators, we are equally pleased that the first rule of problem solving (breaking tough problems down into smaller problems we can solve) 
applies in such a natural way to the task of drawing and animating scenes. 
Not only are children practicing an important
generalizable skill, 
but if they apply this technique recursively,
they generate a hierarchical call tree,
and become primed to learn about simple
higher order functions like $\text{map}$.
They also recognize the importance
of structuring their code so it can be
combined with others' work.

\medskip
But, there is another side of Functional Programming, 
besides functions, namely Algebraic Data Types,
and Elm provides ample motivation for both sum and product types.
The Elm Architecture demands 
the use of a sum type for messages,
and children can relate to this, 
and they readily understand another
sum type:  states in an adventure game.
Most children are just as enthusiastic about
creating interaction as they are about animations.

The one problem in this picture is that the
model-view-update pattern requires the introduction of several new concepts at once: 
the message (sum) type and the update function
containing a case expression.
There is no step-by-step way of introducing them
as solutions to incrementally more complex problems, as with animation.
However, children all have a mental model for adventure games, 
and we had found that they seemed to understand the concept of a state diagram
and its' graphical representation.
This study quantifies this, and gives
us confidence to introduce all children to state diagrams and their representation by sum types.

\subsection{Related Work}

Our background research aimed to identify prior work in the area of using state diagrams to teach
computer science to K-12 students, as well as a more sweeping review of coding education and how it 
relates to other types of literacies.

\subsubsection{Visual Learning and Education}
Based on the keywords used, there seemed to be more literature on learning through drawing than learning through writing in relation to STEM. Ainsworth and Scheiter \cite{ainsworth2021learning} were able to list advantages of drawing: i) limits abstraction ii) exploits “perceptual processes by grouping relevant information” iii) draws on problem-solving instead of memory, iv) provides focus for joint attention/group collaboration, v) increases attention, and vi) activates prior knowledge \cite{chang2012role}. Park et al \cite{park2020sequential} also states that learning through drawings not only takes different perspectives into consideration but also exposes the child to other subject domains (such as math and literacy) when working in groups. Those who used learning-by-drawing scored higher on a test based on comprehension \cite{schmeck2014drawing}. 
Chang \cite{chang2012role} found that when a child and an adult are partners in learning through drawing, that communication was associated with healthy language development and enabled the children to listen, think, and then speak. Interestingly, Cheng and Beal \cite{cheng2020effects} found that while students who drew had a significantly higher cognitive load than those who studied pictures, students were more willing to learn with provided pictures than drawing themselves. That being said, Kunze and Cromley \cite{kunze2021deciding} concluded from their study that ``early secondary'' students are less likely to benefit from ``drawing to learn'' than upper-secondary and post-secondary students.  
So, while visualization can be helpful, whether through examples or by having students create their own visualizations,
the effectiveness of visualization for one age group and subject does not imply effectiveness for other groups, and needs to be evaluated.

\subsubsection{Coding, Literacy, and State Diagrams}
When it comes to child development and coding, looking at coding as a language was heavily emphasized. Again following Piaget, many believe that coding can change the way we think and experience the world around us \cite{bers2019coding}. Coding in itself is seen as a language \cite{bers2019coding}. Goldenberg and Carter \cite{goldenberg2021programming} believe that computer programming is just as important as English and should be taught in elementary school. Monteiro et al \cite{monteiro2021coding} also suggest that ``programming can be the third language that both reduces barriers and provides the missing expressive and creative capabilities children need.'' Coding is a mix of English and math as the words allow for interaction with feedback \cite{goldenberg2021programming}. These two authors also bring up important facts when looking at programming as a language: ``students can construct viable arguments and critique the reasoning with others, it eases the process of beginning with concrete examples and abstracting regularity, perseverance, using the proper tools strategically, and being precise.'' Finally, and akin to English, coding also involves problem-solving, a manipulation of a language, and symbols to create a shareable product \cite{goldenberg2021programming}. 

As discussed previously in \cite{bers2019coding}, previous systems for teaching coding based on Science, Technology, Engineering and Math (STEM) did not account for the intellectual maturation of school-age children. “Coding as Another Language (CAL)” considers coding development alongside language and literacy development. Using similar stages to those of learning reading and writing (emergent, coding and decoding words, fluency, new knowledge, multiple perspective, purposefulness), CAL uses literacy as both a parallel to develop programming curricula and a tool. Knowledge-constructing concept maps can allow for mental mapping of written stories and allow for a newer and easier method for students to record their ideas before starting the writing process according to Anderson-Inman and Horney \cite{anderson1996computer}. State diagrams mimic mind-mapping, a commonly used method of brainstorming in literacy teaching. 
Previous studies and curriculum formed around these standards have shown that it is very possible for students to have a basic understanding of coding upon leaving elementary school \cite{vico2019coding}, similar to their level of reading and writing when entering high school. Recent studies show that lessons in coding can also be useful in teaching mathematics at the elementary level \cite{suters2020coding}. By teaching computational thinking, or the thinking of a computer scientist, at a young age, students are provided with a deeper understanding of mathematical relationships necessary to perform algebra and calculus in later grades. 

\subsection{Using Event-Driving Programming (EDP) in Education}
As Lukkarinen et al. \cite{lukkarinen2021event} state in their literature review of EDP in programming education, event-driven programming and computer programming are two separate entities; programming relies on \textit{organizational charcteristics} whereas EDP focuses on \textit{behavioural characteristics}. For example, while computer programming is more procedural and object oriented, EDP forces the programmer to consider the consequences of the user's actions and how to react to them \cite{lukkarinen2021event}. That being said, the main takeaway is that EDP and computer programming are two different concepts and therefore require two separate ways of teaching. 
On a similar note, they also display the challenges of EDP within their literature review: it is hard to trace the computer programming from beginning to end which affects students abilities to fully understand, EDP has been linked to negative transfer effects when associated with event and non-event oriented programming environments, and ``students who learn to program in an event-fashion do not develop some algorithmic skills that other students will have'' \cite{lukkarinen2021event}. Finally, in terms of challenges, they state that no attempt has been made to alleviate these issues within EDP and learning.

Finally, in this literature review, we learned the most used software tools reported in papers about teaching and learning EDP. They found that Java was the most popular language with App Inventor, C++, and Scratch following far behind. Other tools for Java include DoodlePad and Squint Library \cite{lukkarinen2021event}. There were no functional languages represented, and the authors lament the complete absence of empirical studies comparing different methods of teaching. Methods are not compared to each other, nor to the absence of a strategy for teaching EDP.

Although not advertized as EDP, the Bootstrap\cite{fisler2021evolving} curriculum includes EDP which it describes as a ``modified Model-View-Controller paradigm called Reactors''\footnote{\url{https://bootstrapworld.org/materials/latest/en-us/lessons/re-structures-reactors-animations/index.shtml?pathway=reactive}}.  
Their choice to present interaction this way matches their teaching approach
based on simulation, e.g., a function $\text{state} \to \text{state}$,
which updates the state from one frame to the next in a filmstrip animation.
The two approaches mirror the relationship between 
ordinary differential equations and their closed-form solutions (when they exist).
The deepest understanding comes from understanding both, and the relationship between them.
Our approach treats animation as a closed-form function $\mathbb{R} \rightarrow \text{state}$,
which reinforces the concept of a function,
and has some practical advantages, by elminiating jitter on low-resourced devices unable to draw frames quickly enough, and by allowing ``scrubbing'' (controlling an animation by dragging a slider).

There are other approaches to interaction in functional programming languages,
with different goals.
Many take a pragmatic approach to minimally wrapping 
imperative libraries, 
which can be justified by the fact that
graphical programming is highly motivating to
novice programmers, 
and leveraging advanced concepts will mute this effect. 
For a well-thought-out recent example,
see the Haskell SpriteKit\footnote{\url{http://haskellformac.com/haskell-for-mac-games-in-haskell.html}}.
Others try to simplify the process of
composing graphical ``widgets'' together
graphically, and functionally,
and trace back to at least the Fudgets library
\cite{carlsson1993fudgets},
or to Functional Reactive Animation \cite{Elliott:1997:FRA:258948.258973}.
Note that these approaches make use of visualizations of streams and dataflows
designed to surface the overall structure
of the program which may be difficult to see in the program text.
These approaches were used in the original
Elm run-time system \cite{czaplicki2013asynchronous},
but later abandoned\footnote{http://elm-lang.org/blog/farewell-to-frp} in favour of the reportedly
simpler-to-understand model-view-update
(plus subscriptions which are out of scope for this paper.
Still,
a vibrant community continues to develop functional-reactive 
approaches for new targets.
Perhaps the largest is Reflex,
which is designed by and for knowledgeable 
developers willing to use advanced features
like arrows to obtain elegant representations
of complex systems
at the expense of opacity to beginners (the ``quick reference'' is several 
pages long)\footnote{\url{https://github.com/reflex-frp/reflex/blob/develop/Quickref.md}}.

\subsubsection{State Diagrams in Computer Science Education}
Several authors have used state diagrams to introduce computer science and coding concepts at both the
K-12 and university level. Czejdo and Bhattacharya \cite{czejdo2009programming} discuss lessons using state diagrams
to allow students to describe complex behaviours for robots, which then generated Python code to 
control the robots. They noted that the robots increased the students' engagement with the concepts and
that the diagrams allowed the students to program more complex behaviours than would have been possible
without them.

Kamada \cite{kamada2016islay} similarly used state diagrams to program behaviours of on-screen 
characters, for example programming a virtual fire truck to seek out water and then put out fires. They noted that
``Enthusiastic children often run into the combinatorial explosion of states and transitions. Then 
it is the time for them to move on to the structured programming languages where they can use 
variables to represent states,'' which is aligned with our experience and is discussed later in this paper. They also noted that ``In some disappointing cases, a series of states
are simply chained as if they continue forever'' \cite{kamada2016islay}, which is not something we observed in 
our albeit small study. This, however, 
is continued with the following statement: ``We had better not recommend computer science to those 
children,'' with which we fundamentally disagree. We believe that this should instead be considered a 
teachable moment for the students and an area of improvement for the delivered lesson, instead of jumping to 
the rash conclusion that this indicates a fundamental lack of ability to be a computer scientist.

In contrast to the statement by 
Kamada \cite{kamada2016islay}, Ben-Ari \cite{ben1998constructivism} instead states that
``The science-teaching literature shows that performance
is no indication of understanding. CSE research like
Madison’s, which elicits the internal structures of the
student, is far more helpful than research that measures
performance alone and then draws conclusions on the
success of a technique. A student’s failure to construct a
viable model is a failure of the educational process, even
if the failure is not immediately apparent.'' Thus, in our evaluations we must keep in mind that
failures for students to apply state diagrams should be considered as important lessons for us as 
researchers in how to improve our lessons in the future.

\section{State Diagrams}

The state diagrams of note in this work are a type of non-deterministic finite state automaton (NFA), 
which contains a set of states, transitions and a function mapping from one state to another through
directed transitions. In NFAs, there is no requirement that every state has every possible
transition in the alphabet coming out of it. Unlike the formal definition given in \cite{kozen2012automata}, our diagrams do not have 
final states as they represent programs which continue to respond to user input until closed.

\begin{figure*}[ht]
    \centering
    \includegraphics[width=\textwidth,trim={1cm 0.5cm 0cm 0.5cm}]{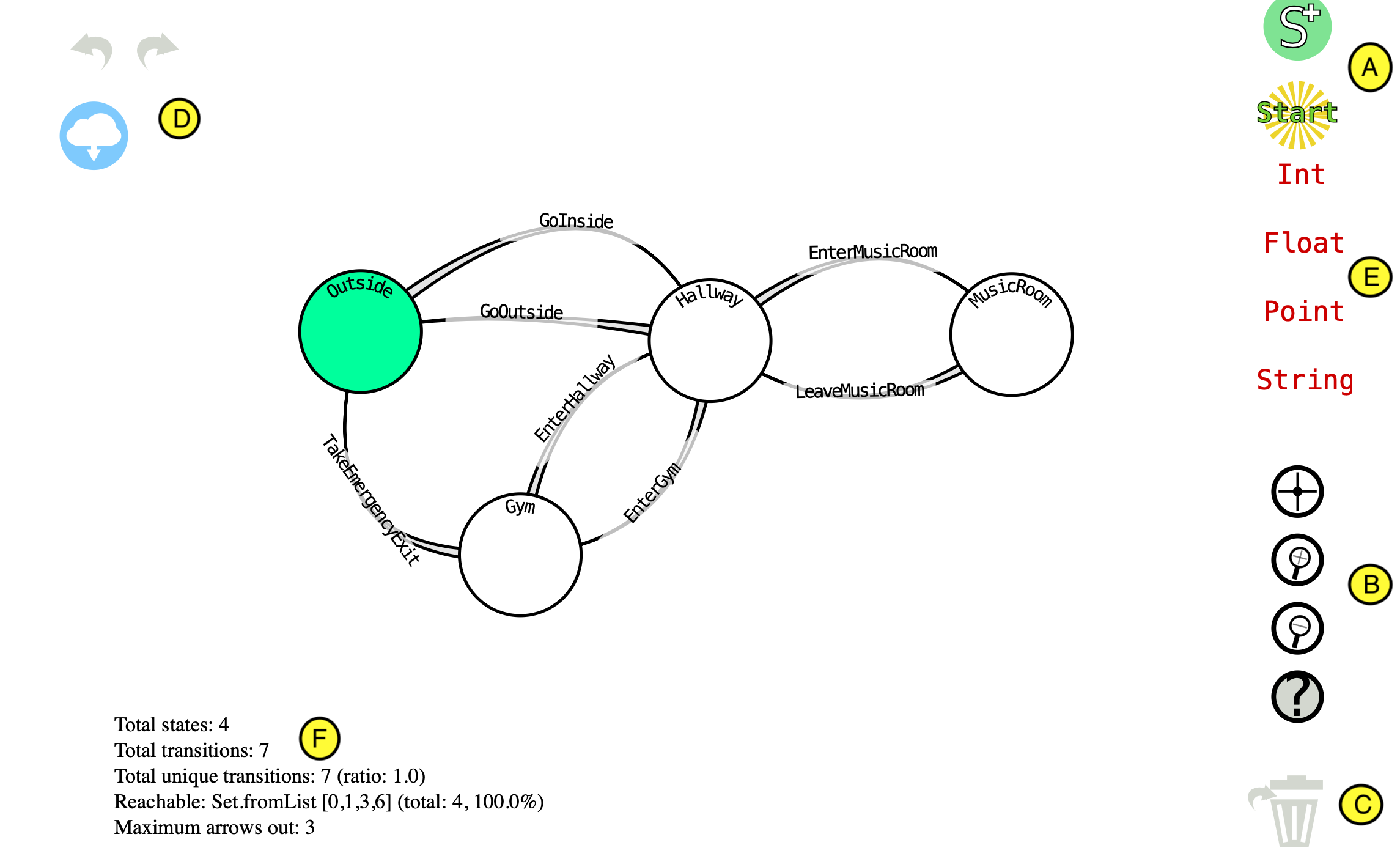}
    \caption{The interface of our web-based state diagram editor, with a diagram representing navigation 
    through a school. (A) allows users to add states or make a state the starting state. (B) has functions for recentering the screen, zooming and a help page. (C) allows states and transitions to be deleted. (D) provides undo / redo and code generation functionality. Although unused in our tests, some built-in types (E) can be added as associated types in messages and states.  To facilitate analysis, we optionally provide summary information (F) on screen.  In the diagram, states are
    represented by circles with arrows (going from wide to narrow) representing transitions from
    one state to another. The green state is the starting state of the diagram.}
    \label{fig:school}
\end{figure*}

\subsection{Graphical Representation and Tool}

A state diagram can be represented graphically using a diagram where states are defined as
labeled circles with transitions drawn as labeled arrows which define legal movements from
state to state. Figure~\ref{fig:school} shows an example state diagram representing the 
navigation of a school.

To facilitate the creation of state diagrams, we have created a tool using the Elm language
with a server backend written using IHP\footnote{https://github.com/digitallyinduced/ihp} in Haskell, 
which currently allows diagrams to be saved to a 
server and accessed later by logging in.
Our state diagram tool allows students to easily draw their state diagrams by defining 
states and then attaching them with transitions. Each state and transition can be given a name
where no pair of states or state and transitions can be named the same. Transitions can be
named the same provided that no state has two transitions with the same name coming from it.
Other invariants needed later for code generation are enforced, such as restricting state names
to only alphanumeric characters starting with a capital letter.

\begin{figure}
    \centering
    \includegraphics[width=0.8\textwidth]{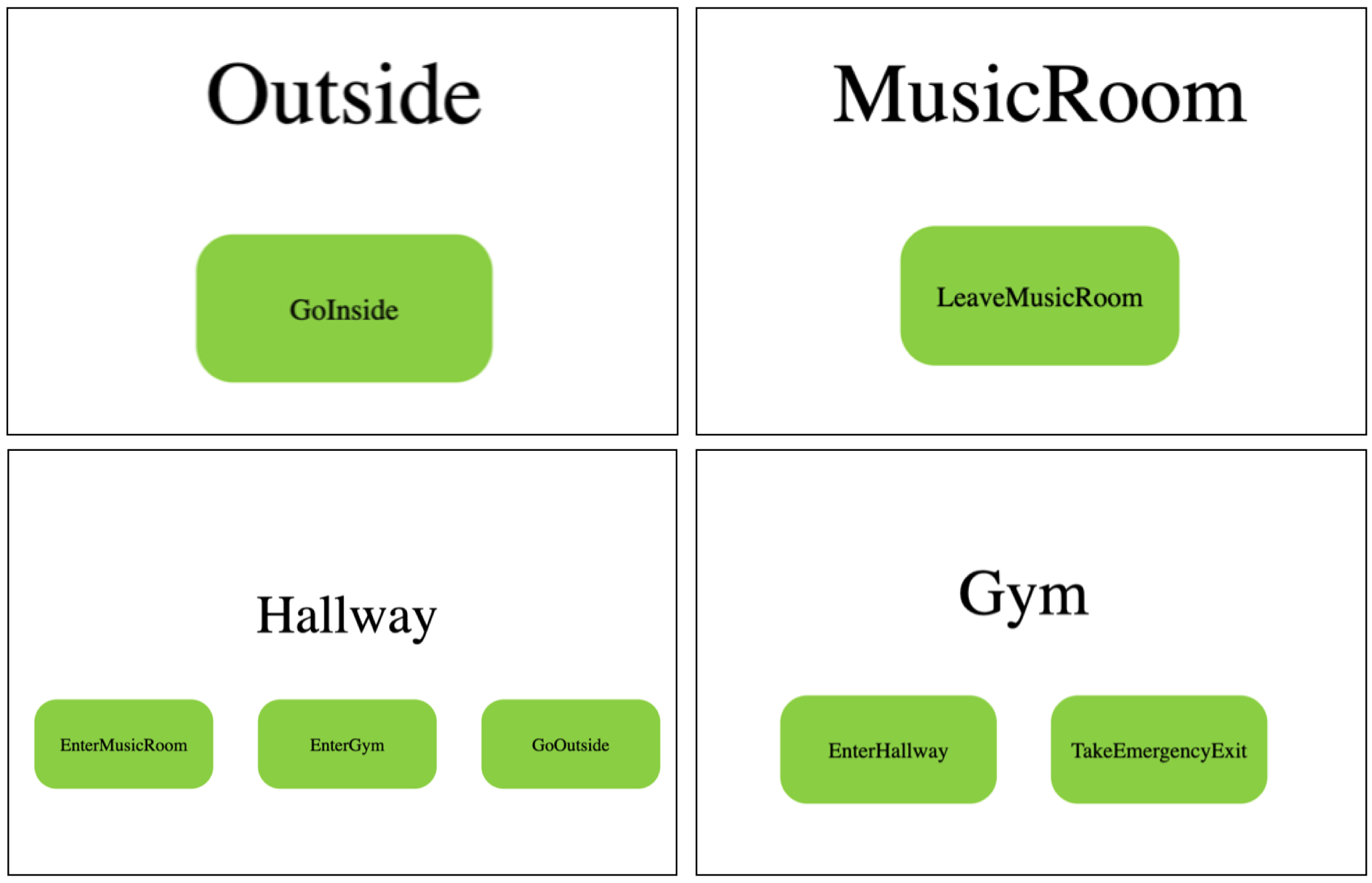}
    \caption{From the state diagram in Figure~\ref{fig:school}, a basic Elm application can be generated 
    using the GraphicSVG library. Shown here are the four different ``pages'' the app can be in, one for 
    each state in the diagram. Each place is given by default a basic title text and buttons for each transition, with the 
    appropriate logic to transition to the correct state when clicked. Students can use existing knowledge
    from previous lessons to code graphics for each page and insert their code into the template into an easy-to-understand case expression.
    In the template, even the buttons are composed from simple shapes so
    learners can easily replace them with themed graphics.}
    \label{fig:exampleApp}
\end{figure}

\subsubsection{Code Generation}

With Elm's algebraic data types, generating the code from a state diagram is straightforward.
The set of possible transitions is mapped to a \texttt{Msg} type and the set of possible states
is mapped to the \texttt{State} type. For instance, for the diagram shown in Figure~\ref{fig:school}, 
the data types generated are respectively:

{\small
\begin{Verbatim}[samepage=true]
type Msg = Tick Float GetKeyState
         | GoInside 
         | EnterMusicRoom 
         | LeaveMusicRoom 
         | EnterGym 
         | EnterHallway 
         | TakeEmergencyExit 
         | GoOutside 
\end{Verbatim}

\begin{Verbatim}[samepage=true]
type State = Outside 
           | Hallway 
           | MusicRoom 
           | Gym 
\end{Verbatim}
}

The \texttt{Tick} message is platform-specific and sends
the current time and keystrokes every 1/30th of a second, for easily making animations and taking 
keyboard input.

The \texttt{update} function is also generated based on the structure of the diagram which includes
the logic for transitioning from state to state. Since all transitions are put into the singule
algebraic data type \texttt{Msg} as constructors, there is no type-level safety for restricting which messages can
be sent from which states. Instead, this logic is generated in the update, which pattern matches
on the current state to determine which state to go to and defaults to keeping the current state
the same if the message is sent from a state where it should not be according to the diagram. There
is, of course, nothing stopping the users from modifying the code so as to no longer match the
diagram, but generally beginners do not need to understand the update function to start creating their
games.

Below is a snippet of the update function generated for the example in Figure~\ref{fig:school}:

{\small
\begin{Verbatim}
update msg model =
    case msg of
        GoInside  ->
            case model.state of
                Outside  ->
                    { model | state = Hallway  }
                otherwise ->
                    model
\end{Verbatim}
}

The \texttt{view} function is generated, pre-populating each place with a basic
text field to identify the current state and basic buttons for transitioning from state to state.
Using knowledge from previous workshops, students can add graphics to each ``page'' of the app
or change the buttons into more interesting objects, such as door handles or levers. See 
Figure~\ref{fig:exampleApp} for an example of how the compiled code looks by default.

\subsubsection{Adding Graphics with Elm}

Once the code is generated, students can compile the code to get a bare-bones app with titles for each
state and buttons representing each transition out of that place. Figure~\ref{fig:exampleApp} shows the
generated app for the example in Figure~\ref{fig:school}. Students can use their existing graphics knowledge 
to create pictures and/or animations for each place.

\section{Design}

This section discusses the design of the lesson and the challenges given to students.

\subsection{Lesson Design}
We developed a lesson plan prior to teaching the Grade 4 students and radically simplified it
when teaching Grade 5 group based on feedback from the classroom teacher.
For Grade 4, we introduced the concept of a state diagram using the Moo-Quack game (see \cite{EPTCS270.2}).
Here, the instructor shows a state diagram in which the states are animal noises and transitions number of 
raised arms. Students are often confused at first but catch on quickly when classmates start making animal 
noises.
We then explained the concepts of a states (places) and transitions (actions) 
using examples drawn in our tool of states of matter, Canadian provincial land borders, and school navigation 
(see Figure~\ref{fig:school}). 
Finally, we showed the children how to generate and run the code in a game slot in our Web IDE.
We then split the class into groups, assigned each to a breakout room, and 
challenged them to make their own game.

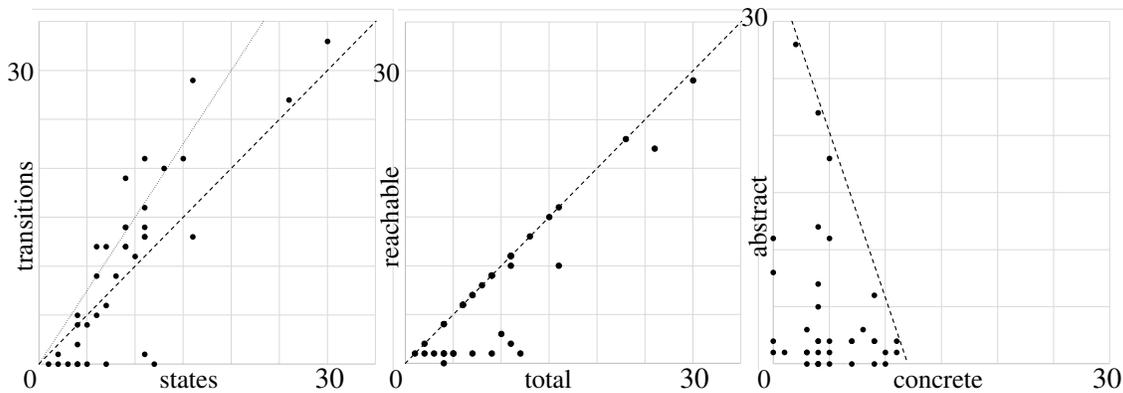
\begin{figure}
    \centering
    \begin{overpic}[trim=40 40 0 0,clip,width=0.3\textwidth]{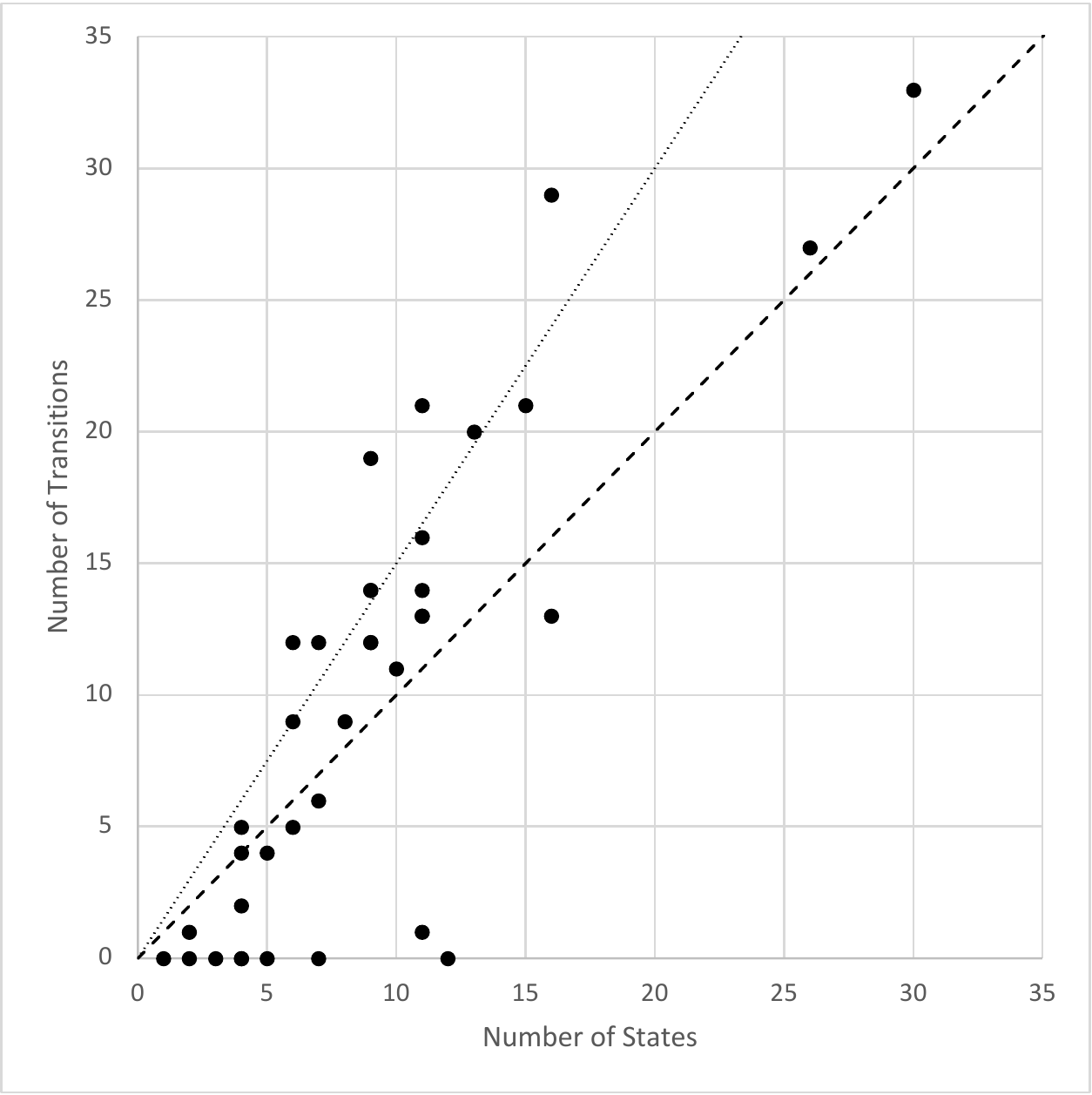}
\put(-2.5,-5){\small{0}}
\put(-7,80){\small{30}}
\put(-5,40){\rotatebox[origin=c]{90}{\small{transitions}}}
\put(35,-5){\small{states}}
\put(78,-5){30}
\end{overpic}
    \begin{overpic}[trim=40 40 0 0,clip,width=0.3\textwidth]{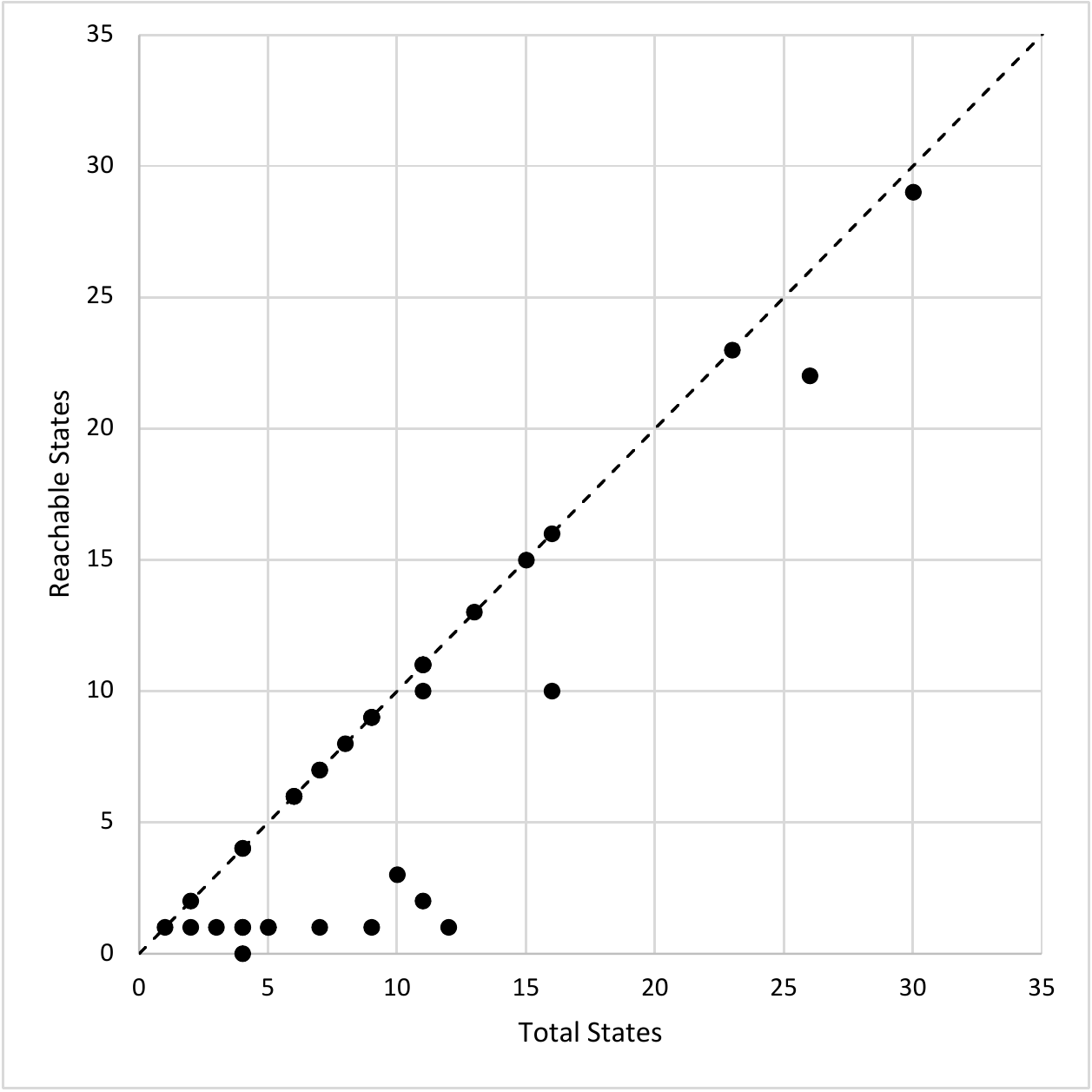}
\put(-2.5,-5){\small{0}}
\put(-7,80){\small{30}}
\put(-5,40){\rotatebox[origin=c]{90}{\small{reachable}}}
\put(35,-5){\small{total}}
\put(78,-5){30}
\end{overpic}
    \begin{overpic}[trim=40 40 0 0,clip,width=0.3\textwidth]{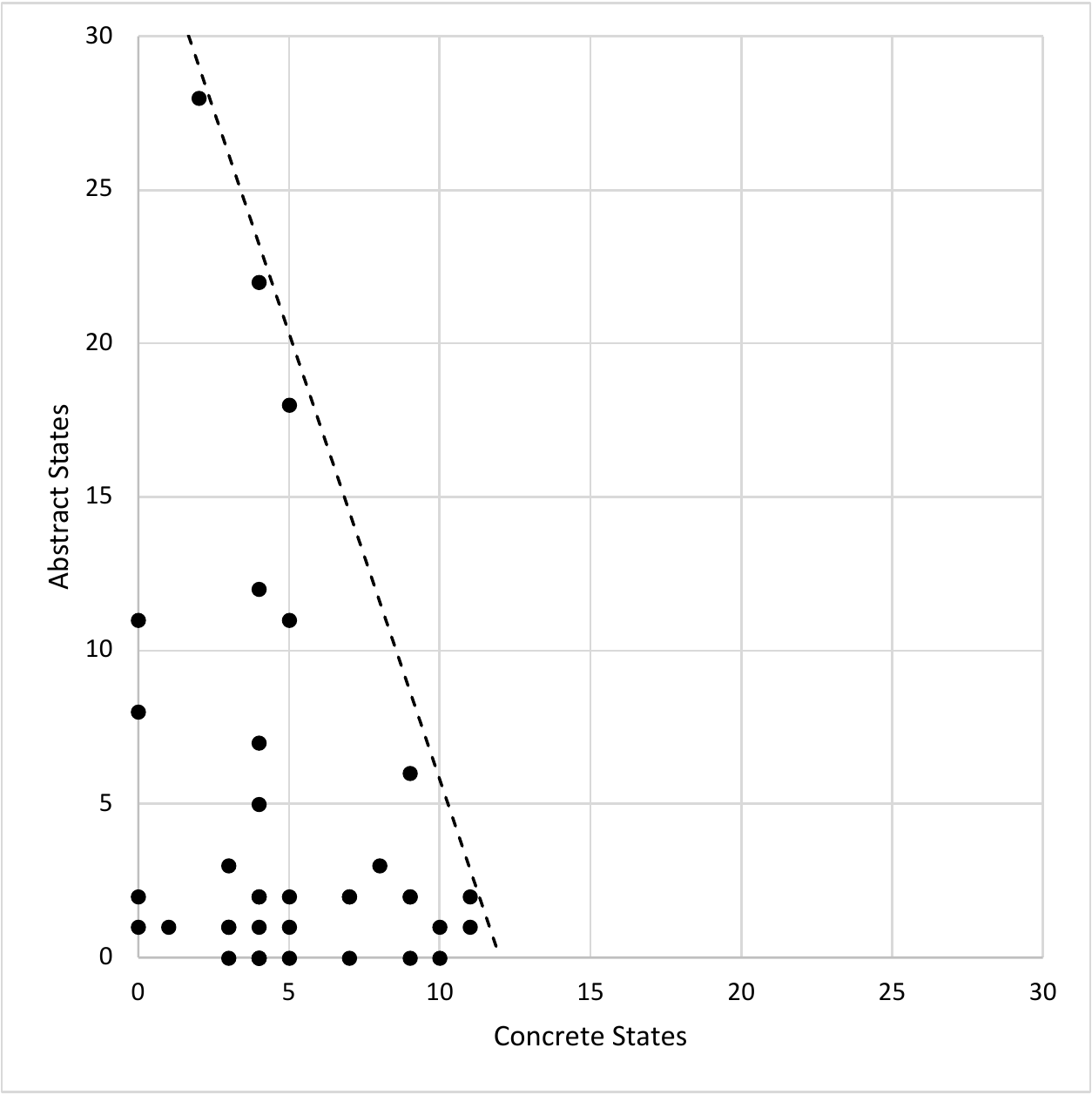}
\put(-2.5,-5){\small{0}}
\put(-7,95){\small{30}}
\put(-5,40){\rotatebox[origin=c]{90}{\small{abstract}}}
\put(35,-5){\small{concrete}}
\put(90,-5){30}
\end{overpic}
    \caption{(top left) A scatter plot of the numbers of transitions and number of states for each diagram. The line $y=x$
            is plotted as a dashed line, and $y=1.5x$ plotted as a dotted line. Points below $y=x$
            indicate more states than transitions and a disconnected graph. The points near the x-axis are likely abandoned diagrams. Points near $y=x$
            indicate diagrams with close to one transition per state, e.g. a tree. Points
            above $y=x$ indicate more complex games with multiple paths.
            (top right) A scatter plot of reachable versus total states in students' diagrams. 
    Points on the diagonal ($y=x$) indicate that all states are reachable from the starting state. The points on the line $y=1$
    probably correspond to abandoned diagrams, since only the starting state is reachable.
    (bottom) A scatter plot of concrete versus states in students' diagrams. 
    The dotted line has slope $-2.9$ which suggests that the effort required to add a concrete state is three times the cost of adding an abstract state.
    }
    \label{fig:trans_div_states}
\end{figure}

We then received feedback from the classroom teachers. For the Grade 5 class, we focused on presenting state diagrams as the 
map of a concrete adventure game and used the vocabulary of ``places'' and ``ways.''
From there, we split the children up into groups and asked each group to chose one person to edit the map while sharing their screen, and we gave them a Google Slides template to use in identifying tasks and then assigning them to group members. 
One of the teachers asked them to think about the scale of their game in terms of 
enjoyment in reading a story.  
We found that if it is too long, people will lose interest.

After class, we were able to retrieve the state diagrams, and approximately assign them to the two grades (i.e., Grade 4 and 5).
Due to the fact our program uses randomly generated logins, we do not keep identifying information, and many of the students continued working on their state diagram after class, we expected to report on the two classes as a whole.
Statistics on the state diagrams are reported below.

\subsection{Challenge Design}
We visited the Grade 5 students the next week for an additional hour 
and gave them four challenges to measure the impact of our teaching
on their understanding of state diagrams and their ability to translate from one representation/implementation to another. Each challenge had two variations; the first one, based on the state diagram in Figure~\ref{fig:school}, contained only concrete place names.
The second variation had abstract states: a closed box with scratching, an open box with a dragon flying around, and a closed box. The challenges were to:
\begin{enumerate}
    \item Draw a state diagram using our tool from an English paragraph:
    \begin{enumerate}[a]
    \item Your task is to make a state diagram to help a new classmate find their way around inside your school. The classmate starts Outside. From the Outside, they can go inside to the Hallway. From the Hallway, they can enter the Music Room. From the Music Room they can leave and go back into the Hallway. From the Hallway they can also enter the Gym. From the Gym they can leave the Gym and go back to the Hallway, or in an emergency they can take the emergency exit to go back Outside. Your classmate cannot enter the school through the emergency exit.
    \item You are designing a state diagram for a video game about a dragon. The game starts with a Closed Chest, with a scratching noise inside. The player can open the chest, which will cause a dragon to start flying around. The player can then close and open the chest as many times as they want, but the dragon will still be flying around and the chest will remain empty. 
    \end{enumerate}
    \item Draw a state diagram using our tool from bullet points. (See \cite{pasupathi2021sd} for lists of the points.)
    \item Describe a state diagram using English based on a diagram drawing using our tool.
    \item Draw a state diagram using our tool given a generated (and compiled) game, e.g. Figure~\ref{fig:exampleApp}.
\end{enumerate}

\section{Results}
The detailed result analysis of the above challenges are discussed here. 
The median results are highlighted here. Students who learned state diagrams based on adventure game came up with better diagrams than the other students. The statistical analysis are based on some of the factors like number of states and transitions used in a diagram, abstract names and concrete names for states, and reachability of states through transitions. 

\subsection{Quantitative Analysis of State Diagrams}
\label{sec:quantitativeAnalysis}

Figure~\ref{fig:trans_div_states} shows scatter plots of (left) the total number of transitions versus states, 
(middle) reachable vs total states, and 
(right) abstract versus concrete states.
Consistent with our observation that many diagrams were abandoned before adding transitions,
we note a cluster of disconnected diagrams,
but otherwise the diagrams indicate strong understanding of and engagement with the material. 
There was widespread adoption of abstract states, even though most students were only taught to think about and create concrete states.
In fact, most productive groups produced many more abstract than concrete states.

\begin{figure}
    \centering
    \includegraphics[width=0.8\textwidth]{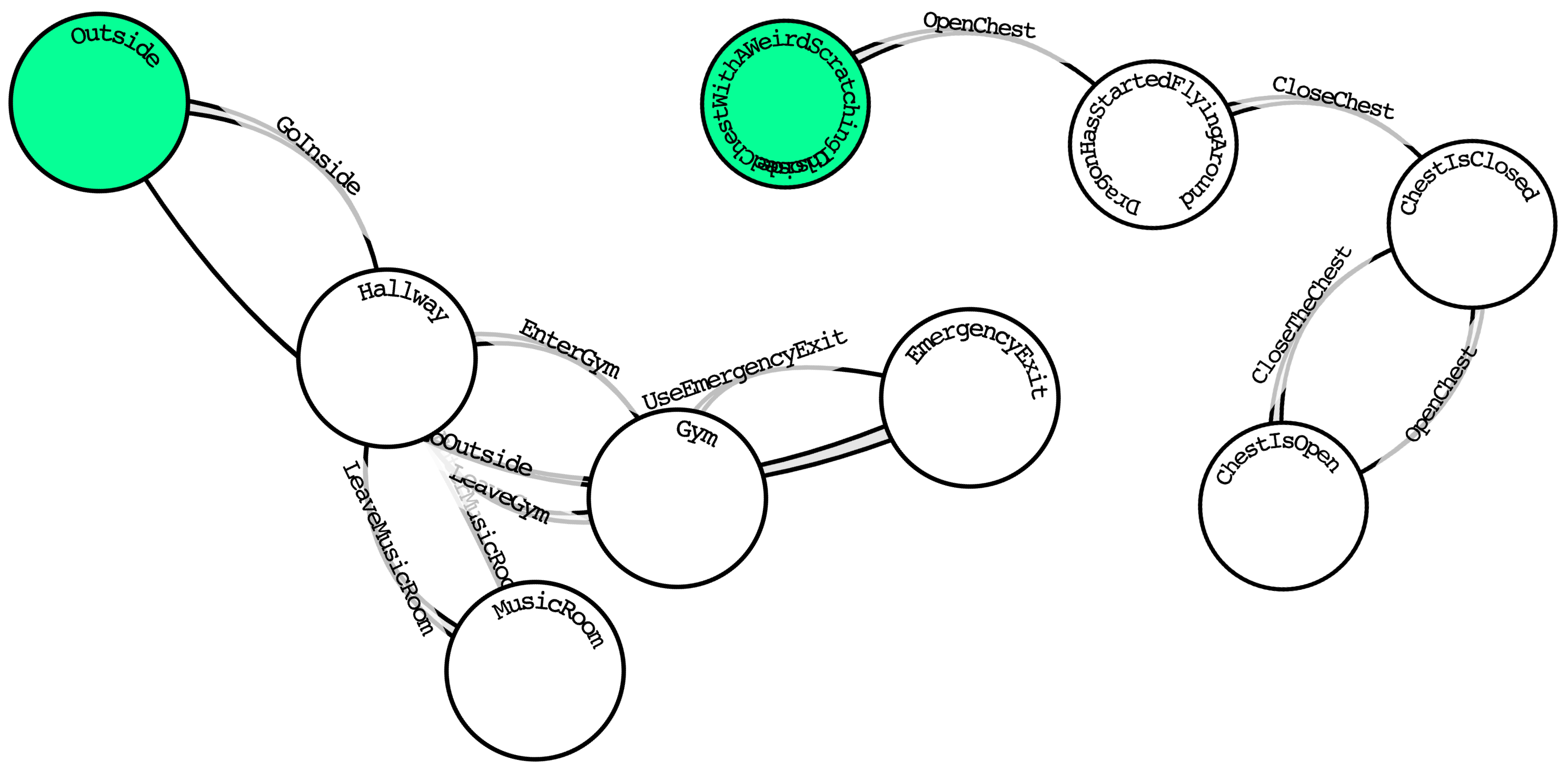}
    \caption{Median results for challenges 1a (left) and 1b (right).  Note that there is no hard limit on the state names, and they can overlap in the diagram, although they generate working code.}
    \label{fig:challenge11}
\end{figure}

\medskip

To answer \textbf{RQ3},
we have tested the hypothesis that the diagrams created come from the distribution of randomly generated diagrams,
generated from the uniform distribution on the set of subsets of possible directed edges with a given cardinality.
To use the Anderson-Darling single-sample test, we need to know the cumulative probability distribution (CDF) for the number of states reachable from the starting state.
We approximated this discrete distribution by generating random diagrams.
Using the empirical CDF, we can randomly generate samples of diagrams and evaluate the Anderson-Darling statistic to approximate its distribution,
and estimate the $p$-value for the child-created diagrams.

\begin{table}[!ht]
    \centering
    \begin{tabular}{|l|l|p{7cm}|l|}
    \hline
         States & Transitions & Probability Distribution Function  & Observed \\
    \hline 11 & 13 & \includegraphics[width=7cm]{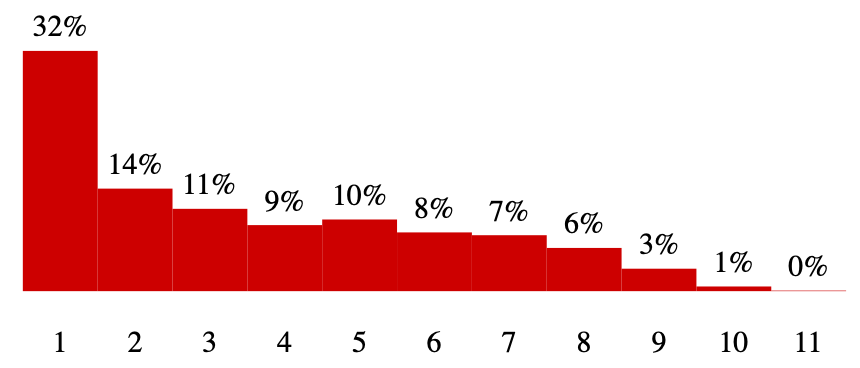} & 10,11\\
    \hline 11 & 14 & \includegraphics[width=7cm]{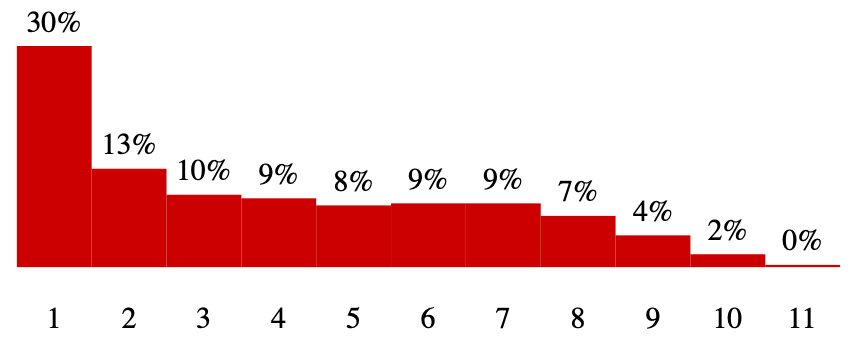} & 11\\
    \hline 11 & 16 & \includegraphics[width=7cm]{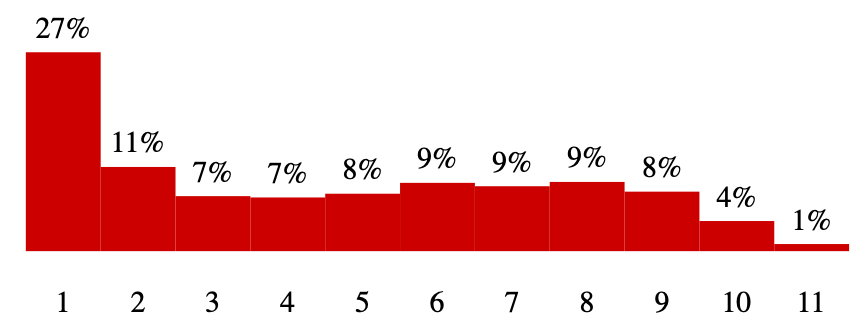} & 11\\
    \hline 11 & 21 & \includegraphics[width=7cm]{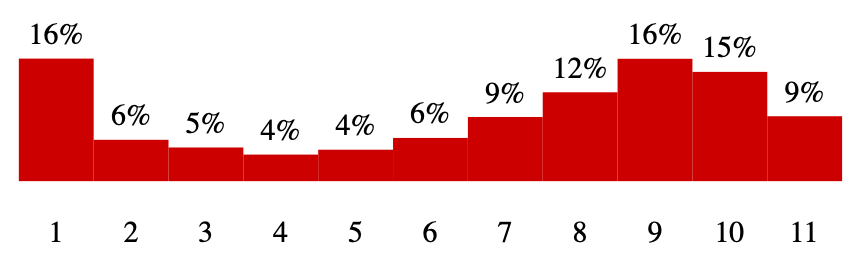} & 11\\
    \hline
    \end{tabular}
    \caption{Probability Distribution functions for the number of reachable states based on simulation of 4000 random diagrams with 11 states. The observed column has the number of reachable states in the students' state diagrams with 11 states and its corresponding transitions. See \cite{pasupathi2021sd} for more.}
    \label{tab:PDF1}
\end{table}

\smallskip
In Table~\ref{tab:PDF1},
we display the probability distribution for reachability for each diagram with respect to the number of states and transitions in the diagram, for the case of diagrams with 11 states, as well as the observed rechabilities.
Since the initial state is always reachable, the minimum reachability is one.

The Anderson-Darling test statistic measures the match between a sample and a distribution. 
We computed compared the observed reachabilities against randomly generated directed graphs with 11 states, and found that with confidence $p<0.001$, that these four diagrams are not a sampling of the distribution of uniform random graphs.
See \cite{pasupathi2021sd} for additional cases, and detailed calculation of the Anderson-Darling test statistic.

\subsection{Qualitative Analysis of Challenges}
\label{sec:qualitativeAnalysis}

We report the results of the challenges qualitatively because
(1) to understand where we needed to improve instruction without asking for too much student time we had 3-5 students completing each challenge,
(2) we asked some children to write paragraphs, and
(3) we asked some children to create state diagrams.
In all cases, we could identify a ``median'' response, which we report here.

\begin{figure}
    \centering
    \includegraphics[width=0.99\textwidth]{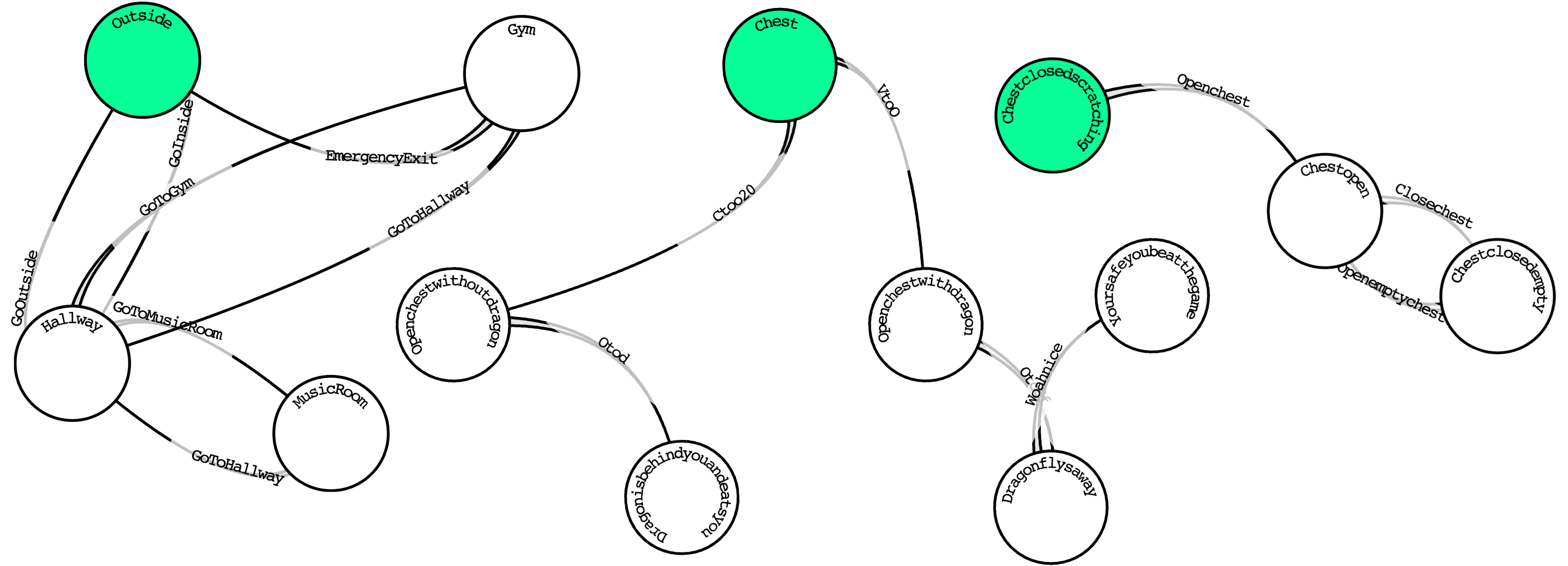}
    \caption{Median results for challenges 2a (left), 2b (middle) and 4b (right).}
    \label{fig:challenge21}
\end{figure}

In the concrete school challenges, we found confusion about whether
the emergency exit was a state or a transition, as seen in Figure~\ref{fig:challenge11} when translating from a paragraph description.
Whereas given a point-form description, they were less likely to leave off or add additional states or transitions, as seen in Figure~\ref{fig:challenge21}.

When asked to describe the state diagram in English, 
most students hesitated to get started, but then used narrative to thread together a description.  Many students asked ``when should I stop?'' because they realized that the narrative could go on forever due to a cycle, giving a glimpse into how they were systematically analyzing the diagrams.
Our median response for Challenge 3a was: 
\textit{You start outside the school. If you go inside through the door, you'll be at the hallway. Here, you can access all the different rooms or exit the hallway to go back outside. The music room is the room labeled ``music,'' and you can enter and exit it through the door. From the hallway, you can also enter the gym room, and exit it back through the door. If you are in the gym and there is an emergency, you can take the emergency exit instead of running back to the hallway and exiting that way. There is no emergency exit in the music room, since there is nowhere to go after you leave.}
There was no median response to the final challenge due to miscommunication about the need to work individually.

The abstract dragon-in-a-box challenge was more difficult, but they did best when translating a working app into a diagram, see Figure~\ref{fig:challenge21},
perhaps because they did not have to think about the semantics of the abstract situation.  When they had to reason about the English description, they were uncertain, but still seemed to understand the basic definitions and task, see Figures~\ref{fig:challenge11} and \ref{fig:challenge21}.

Going the other way,
they again showed their understanding of the concept, but felt the need to add narrative bridges, perhaps because we used narrative in our descriptions, 
or because that is how they understand them.  Our median solution:
\textit{You start with a chest, and you hear a faint scratching noise inside. If you run away, you will find that the room has no escape, and you have to open the chest. When you open the chest, The dragon flies out, and all you have to do is to close the chest, and you win.}
The inclusion of narrative elements to explain properties of the state diagram was surprising and shows their understanding of the state diagram model. In this response, \textit{you will find that the room has no escape} shows that they understand how the diagram models the allowable transitions at a given place. The addition of \textit{and you win} shows another common observation we saw with their responses: the student adding their pre-existing knowledge of video games to their responses, whether or not that information is encoded in the diagram, which in this case it is not.

\section{Discussion of Results}
The primary purpose of this study was to test the tool and pedagogy of teaching state diagrams. Identified future work is presented in this section.

One of the more interesting results of the classes was that the students began to differentiate between concrete states (places one could go, e.g., \texttt{Hallway}) and abstract states (new states of being, e.g., \texttt{DragonHasStartedFlyingAround}), and apply different levels of both. For example, in a game where a character can enter a barn, he finds a dragon. The character can transition to a new state by feeding the dragon. The barn itself would be a concrete state, as it is a place the character can enter. The fed dragon would be an abstract state, as you cannot return to the state of the dragon being unfed. Both types of states were given in the examples provided in class; however, it was never explained that there might be a difference between the two. 

Once the results of what the students had created was reviewed, it was hypothesized that they could be sorted by complexity in two forms: number of states and transitions, and number of concrete and abstract transitions. The results showed few games of moderate complexity, instead showing that students favoured either high complexity or low complexity. Students who used more abstract states also had higher rates of states and transitions, while those with more concrete states had less states and transitions. Further research might be able to explain why this trend was seen. 

It was also noted, when checking in on the state diagrams created by the students, that some continued their work after the classes had ended. Though these results were not analyzed due to the small participation number, it seemed that those who showed more complexity by using abstract states and a higher number of states and transitions overall, were the same students who continued working on their projects in their own time. This showed the investment the students had in their stories that had not been predicted. As mentioned previously, the ability to visually map out ideas through concept mapping or state diagrams has been shown to improve the efficiency of a student's writing. However, this finding also suggests the potential for higher engagement, interest and initiative when learning to code, and create culminating projects fusing coding and other curriculum areas.

\subsection{Limitations}

Firstly, our results are based on approximately 38 diagrams from a total of 70 students from an enrichment program. (Children worked in small groups.)  
Secondly, we did not follow the same lesson plan when teaching the Grade 4s as when teaching the Grade 5s. After one lesson with the Grade 4 students, a streamlined teaching plan was used for the Grade 5 students. Because we use anonymized accounts, we cannot segregate and compare their results. 
Thirdly, it is unlikely that random graphs drawn by a child would come from the uniform distribution.
Even if a child does not understand the diagram as a state diagram,
they may still understand it in some other way and choose random edges from some other aesthetic,
but figuring out such a distribution would be very difficult, 
and it is as posible that such distributions
would be \textit{less} connected than ours.
For example, a child might arrange vertices as petals on multiple flowers, with each petal being a dead-end, and no way to get to the centre.
The uniform distribution is the simplest 
distribution which fits our limited knowledge of the case, 
and we can have some confidence because the observed distribution is so very unlikely given the uniformly random hypothesis. 
This also matches the impression of the instructors that the children were able to discuss the game maps, and spontaneously introduce the concept of a ``trap door'' in their discussions. 
Finally, we could not control for the student's device and internet connection.  

\section{Conclusions and Future Work}

Our initial motivations for this work were (1) the positive reaction of children to our previous tool, PAL~Draw, and (2) questions raised after our TFPIE 2017 presentation about the advisability of discussing state with novice functional programmers.
We are satisfied that this proof-of-concept is ready for universal use in our own outreach program,
and we hope that positive answers to our research questions (below) will help convince other educators of the wisdom of using graphical state diagrams to teach interaction to children learning a functional language.
\begin{enumerate}
    \item[RQ1] \textit{Do grade 4-5 students demonstrate an understanding of State Diagrams by being able to translate between different representations?} Yes, when students were given different representations and asked to convert them, they were able to do so. See Section~\ref{sec:qualitativeAnalysis}.
    \item[RQ2] \textit{Do grade 4-5 students demonstrate equal facility for translating between different representations of state diagrams?}   
    No. When a state diagram is given and students are asked to write a description about it, students were confused how much to write,
    especially in the case of a cycle in the state diagram.
    Students also found it easier to interpret point-form specifications rather than paragraphs,
    and found the conversion of a working app into a State Diagram easiest of all. See Section~\ref{sec:qualitativeAnalysis}.

    \item[RQ3] \textit{Can grade 4-5 students understand the role of reachability? Assuming that students who did not understand the role of reachability would generate random graphs, what confidence do we have that the graphs are more reachable than random graphs?}  We have a confidence of $p<0.001$ that they understand reachability enough not to draw random diagrams, based on an monte carlo simulation of the Anderson-Darling statistic for four of the diagrams with 11 states. See Section~\ref{sec:quantitativeAnalysis}.
    
    \item[RQ4] \textit{Are grade 4-5 students engaged by state diagrams and their applications to adventure games?} Yes, students were engaged by state diagrams. They expressed interest in creating the state diagrams during class, and showed even more interest in designing the levels or difficulty of the levels of their game than designing the graphics for their game. Several groups continued working after school.
    
    \item[RQ5] \textit{Do grade 4-5 students understand abstract and concrete states equally well? Will students presented with concrete states generalize to abstract states without prompting?} For the grade 5 students, we did not explain about concrete and abstract states. But when we asked them to draw their own State Diagrams for their favorite game, some students came up with abstract states. See Figure~\ref{fig:trans_div_states} (bottom).
\end{enumerate}

\noindent
That said, we have plans for tool and pedagoical improvements and future studies.

\subsection{Tool Improvements}

Given that most of our teaching has been forced online due to the ongoing COVID-19 pandemic, more features
are planned for improving distance education. 
Instructors can currently view and make changes to 
students' state diagrams,
but live editing and viewing and shared control by teams would significantly 
improve distance learning and teamwork.

The next step towards model-driven engineering requires
the integration of the state diagram editor with 
our Web IDE, allowing the code generation button to automatically 
open a game slot with the generated code.
Full model-driven development would add the ability to make
changes in the state diagram and have them mirrored in the code.
This is more work, but is important to support an iterative design thinking approach to development.

Collaboration could also be supported by allowing sub- or nested- state diagrams; that is, allowing an entire state
diagram to exist within a state of a larger diagram. 
Students are interested in designing
mini-games as part of a larger game,
and in fact we encourage this with a summer camp. 
Nested state diagrams would allow students to integrate mini-games
without mentor involvement, as is currently required
\footnote{\url{http://outreach.mcmaster.ca/\#camps} and \url{https://macoutreach.rocks/escapemathisland/}}.
This would require that sub-diagrams can be tested independently,
similar to the support we already provide for individual frames in animated comics.

Beginner students can get very far with data-less states and transitions but eventually fall prey to what is
known as a ``state explosion,'' where students create many states and transitions to represent data which 
would make more sense as a data type like a Boolean, integer, etc. 
Even in our first 2.5-hour workshop 
we saw students making such diagrams, with states representing things like the amount of health an enemy
has left and transitions for dealing with different damage values. We believe that students do benefit from the 
discussion and problem-solving that went into making such complex diagrams, based on discussions we overheard during the session. Furthermore, generally students are not 
ready for things like integers being added to their states and transitions until they have had the chance to 
design their diagrams, generate the code, and discover the state explosion problem on their own. However, 
especially (or perhaps, only) when these tools are used in longer-term settings like a summer camp, eventually 
the basic ``untyped'' diagram is no longer powerful enough to support the students' ideas. Thus, future work 
includes finding the best way to introduce and teach these concepts, as well as support for user-defined algebraic 
data types and an interface to model such data. 

\subsection{Pedagogical Improvements}

We found that students were generally able to translate between different
equivalent specifications of state diagrams,
but that they were more successful and (anecdotally) more comfortable with point-form specifications than paragraphs.   
This suggests designing and testing a staged curriculum in which 
translation between diagrams and point-form specifications should be taught first,
followed by paragraph descriptions,
and the advantages of translating from working game to diagram to point-form to paragraph should be measured.
Teachers are always trying to find ways of engaging reluctant writers,
and we hypothesize that this is one way of leveraging children's engagement with video games.
Different teaching styles should be investigated for different age groups.  
Furthermore, demographic information collected beforehand can allow the foundation for comparing genders, ages and developmental differences. 

Knowing that many students created abstract states without prompting, see Figure~\ref{fig:trans_div_states},
we should add a follow-up lesson after children have produced one (or more) state diagrams to introduce abstract states to all children.
We should then design new challenges to evaluate whether all children
are able to understand and use abstract states,
and whether there are differences based on developmental level.

\subsection{Possible Impact on Functional Programming Education}

Anecdotally, we often see undergraduate learners with a lot of experience in other languages add new fields to the Model record type,
including Boolean variables indicating when other variables are valid,
sometimes Maybe variables. 
Unfortunately, they never use sum types.
We would argue that the use of sum types in refactoring data models to make them safe-by-design should be emphasized in functional programming courses.
SD~Draw with code generation provides a concrete example of sum types,
and we have observed first-year students using it to good effect.

It is well known that undergraduates have faulty mental models of recursion \cite{mccauley2015teaching},
and that having a mental model of the function stack is among the strongest indicators that their mental models of recursion are sound.
Unfortunately, the function stack is a run-time artifact, 
and not something they can see in the code without a lot of scrolling back and forth, which increases cognitive load unhelpfully.
On the other hand, a recursive data type in Elm, or even a collection of mutually recursive data types, have compact representations which are easy to see.
Moreover, unlike the function stack which needs to be understood as dynamic,
recursive data types used for information retrieval can be used (and understood) as static objects.

Although this idea should be formulated and experimentally tested,
anecdotally, we have received feedback from upper-year undergraduate students who report writing prototype solutions in Haskell as their preferred way to do data modeling, followed by translating into the language required for the assignment.
In other words, they prefer writing (and compiling) partial Haskell programs to document the structure of their programs over using ad-hoc diagrams, UML or pseudo-code.

\medskip
Why is this approach not a mainstay of programming courses?
It may be because influential textbooks favoured other strategies:
either a focus on substition as an operational semantics for 
programs, which we attribute to Bird \cite{bird1987introduction},
or a tradition of teaching using (untyped) Lisp (and later Scheme and Racket) whose best expression is
\textit{How to Design Programs} \cite{felleisen2018design}.
Since lambda calculus was originally developed without types, 
approaches build around it are unlikely to  teach 
types as an essential tool for beginners trying to understand how computation works.
In the Lisp school, even Racket does not have a clean syntax for sum types,
so using them as a tool for understanding means learning has to compete for cognitive resources with code parsing.

\vspace{-0.25cm}

\section*{Acknowledgements}

\vspace{-0.25cm}

We acknowledge financial support from the Faculty of Engineering, and moral support of the teachers, parents, and young coders.

\vspace{-0.25cm}


\end{document}